\def\polylog{\mathop{\rm polylog}}
\def\fq{\mathbb{F}_q}
\def\fp{\mathbb{F}_p}
\def\deg{\mathop{\rm deg}}

\def\min{\mathop{\rm min}}
\def\max{\mathop{\rm max}}
\def\wgt{\mathop{\rm wgt}}

\def\ord{\mathop{\rm ord}}
\def\css{\mathop{\rm CSS}\nolimits}
\def\gb{\mathop{\rm GB}\nolimits}

\documentclass[reprint,amsmath,amssymb,pra,aps,
longbibliography,tightenlines,notitlepage]{revtex4-2}
\usepackage{graphicx}
\graphicspath{{.}{./figs/}}

\usepackage{dcolumn}
\usepackage{bm}
\usepackage[colorlinks,citecolor=blue]{hyperref}
\usepackage{amsthm}
\usepackage{physics}
\usepackage{csquotes}            

\usepackage[inline]{enumitem}
\setlist{nosep}

\usepackage{csquotes}            

\newtheorem{theorem}{Theorem}

\newtheorem{statement}[theorem]{Statement}
\newtheorem{corollary}[theorem]{Corollary}

\newtheorem{example}[theorem]{Example}

\begin{document}

\title{Distance bounds for generalized bicycle codes}

\author{Renyu Wang}

\affiliation{%
  Department of Physics \& Astronomy, University of California,
  Riverside, California 92521, USA }

\author{Leonid P. Pryadko}

\affiliation{%
  Department of Physics \& Astronomy, University of California,
  Riverside, California 92521, USA }%

\date\today

\begin{abstract}
  Generalized bicycle (GB) codes is a class of quantum
  error-correcting codes constructed from a pair of binary circulant
  matrices.  Unlike for other simple quantum code ans\"atze,
  unrestricted GB codes may have linear distance scaling.  In
  addition, low-density parity-check GB codes have a naturally
  overcomplete set of low-weight stabilizer generators, which is
  expected to improve their performance in the presence of syndrome
  measurement errors.  For such GB codes with a given maximum
  generator weight $w$, we constructed upper distance bounds by
  mapping them to codes local in $D\le w-1$ dimensions, and lower
  existence bounds which give $d\ge {\cal O}({n}^{1/2})$.  We have also
  done an exhaustive enumeration of GB codes for certain prime
  circulant sizes in a family of two-qubit encoding codes with row
  weights 4, 6, and 8; the observed distance scaling is consistent
  with $A(w){n}^{1/2}+B(w)$, where $n$ is the code length and $A(w)$ is
  increasing with $w$.
\end{abstract}

\maketitle


\section{Introduction}
\label{sec:intro}

In the last two years there was an enormous progress in the theory of
quantum low-density parity-check (LDPC)
codes\cite{Evra-Kaufman-Zemor-2020,%
  Hastings-Haah-ODonnell-2020,Panteleev-Kalachev-2020,%
  Breuckmann-Eberhardt-2020,Panteleev-Kalachev-2021}.  Such code
families, with bounded weight of stabilizer generators and distance
scaling logarithmically or faster with the block length, generally
have a finite threshold to scalable error
correction\cite{Dennis-Kitaev-Landahl-Preskill-2002,%
  Kovalev-Pryadko-FT-2013,Dumer-Kovalev-Pryadko-bnd-2015}.  Unlike in
the case of classical LDPC codes\cite{Gallager-book-1963,%
  Chung-Forney-Richardson-Urbanke-2001} where sparse random matrices
can be used to define the code, due to a commutativity constraint, an
algebraic ansatz is required in the case of quantum LDPC codes.  For
over a decade, no construction was known to give distances larger than
a square root of the block size $n$, up to a polylogarithmic
factor\cite{kitaev-anyons,Dennis-Kitaev-Landahl-Preskill-2002,%
  Freedman-Meyer-Luo-2002, Tillich-Zemor-2009,
  Kovalev-Pryadko-Hyperbicycle-2013,
  Guth-Lubotzky-2014,Evra-Kaufman-Zemor-2020,Zeng-Pryadko-2018,%
  Zeng-Pryadko-hprod-2020,Kaufman-Tessler-2021}.  The
$\mathcal{O}\biglb(\sqrt{n}\polylog n\bigrb)$ barrier was broken by
Hastings, Haah, and O'Donnell\cite{Hastings-Haah-ODonnell-2020} who
demonstrated a code family with the distance
$\mathcal{O}(n^{3/5}/\polylog n)$.  Soon followed related
constructions\cite{Panteleev-Kalachev-2020,%
  Breuckmann-Eberhardt-2020}, with Panteleev and
Kalachev\cite{Panteleev-Kalachev-2021} finally proving the existence
of asymptotically good bounded-stabilizer-generator-weight LDPC codes,
with both the asymptotic rate and the asymptotic relative
distance non-zero.\\

Unfortunately, the constructions in
Refs.~\onlinecite{Evra-Kaufman-Zemor-2020,Hastings-Haah-ODonnell-2020,%
  Panteleev-Kalachev-2020,Breuckmann-Eberhardt-2020,%
  Panteleev-Kalachev-2021} do not come with an estimate for stabilizer
generator weights sufficient for getting good quantum codes, or if
they do, not one small enough to give practical codes.  Further, these
anz\"atse tend to produce rather long codes; shorter codes obtained
this way may
have parameters not as good as with constructions known earlier.\\

In comparison, generalized bicycle (GB)
codes\cite{Kovalev-Pryadko-Hyperbicycle-2013,%
  Panteleev-Kalachev-2019}, a generalization of the bicycle
construction from Ref.~\onlinecite{MacKay-Mitchison-McFadden-2004},
are particularly suited for constructing short codes, as a GB code can
be constructed from a pair of linear cyclic codes which are only a
factor of two shorter.  Second, as we show in this work, a subset of
codes from several well-studied families, most notably, quantum
hypergraph-product (QHP) codes in two and higher
dimensions\cite{Tillich-Zemor-2009,Zeng-Pryadko-2018,
  Zeng-Pryadko-hprod-2020}, including the codes with finite asymptotic
rates and power-law distance scaling, can be mapped to bicycle codes.
At the same time, the distance bound $d\le n^{1/2}$ which limits the
parameters of all QHP codes, does not apply to GB codes; we show in
this work that this family includes codes with linear distances.
Third, regular structure of GB codes simplifies both their
implementation and linear-complexity iterative
decoding\cite{Panteleev-Kalachev-2019,Raveendran-Vasic-2021}.
Moreover, GB codes have naturally overcomplete sets of minimum-weight
stabilizer generators, which may improve their performance in the
fault-tolerant (FT) setting.  In spite of these advantages and the
long history of GB codes, their properties have not been
systematically studied.

The goal of this work is to investigate the parameters of GB codes,
targeting highly-degenerate codes with distances much larger than the
stabilizer generator weight which for practical codes should stay
under $w_{\rm max}\simeq 10$.  While some of the present distance
bounds are an easy consequence of those obtained for related codes, or
are obtained with well known methods, we believe a systematic review
of available results is necessary.  These results include
Gilbert-Varshamov-style existence bounds for unrestricted GB codes,
upper bounds for parameters of GB codes with row weight $w$ obtained
by a map to codes local in $D\le w-1$ dimensions, and several expicit
constructions.  Other results include an exact expression for the
distance in terms of an associated asymmetric quantum code, a matching
set of upper and lower distance bounds for $w=4$ bicycle codes, and a
lower bound which guarantees the existence of long GB codes with the
distance $\mathcal{O}(n^{1/2})$ for any fixed $w\ge 4$.  We also
studied the family of GB codes known to include codes with linear
distances numerically, by exhaustively enumerating the corresponding
binary GB codes with row weights $w=4$, $6$ and $8$, for circulant
sizes $\ell\le 217$ with primitive root 2.  Although we are not able
to distinguish conclusively between a power-law distance scaling
$d=\mathcal{O}(n^\alpha)$ with $\alpha=1/2$ and $\alpha>1/2$, the
results are consistent with square root distance scaling and a
prefactor an increasing function of $w$.

The structure of the paper is as follows.  First, in
Sec.~\ref{sec:review} we give a brief summary of relevant facts from
the theory of classical and quantum error-correcting codes, including
some information on cyclic and quasi-cyclic codes.  Analytical results
are collected in Sec.~\ref{sec:GB-codes}.  Namely,
Sec.~\ref{sec:GB-def} gives general information about GB codes,
Sec.~\ref{sec:GV-bounds} collects several lower (existence) bounds on
distances of unrestricted GB codes based on the CSS map,
Sec.~\ref{sec:QHP-map} gives existence bounds based on the map to
hypergraph-product and related codes, Sec.~\ref{sec:D-map} gives a map
of a weight-$w$ GB code to a code local in $D\le w-1$ dimensions, and
Sec.~\ref{sec:weight-four} gives tight bounds for weight-four GB
codes.  Numerical results are collected in Sec.~\ref{sec:numeric},
followed by a brief Conclusion in Sec.~\ref{sec:conclusion}.  Some of
the formal proofs are collected in the Appendix \ref{sec:appendix}.

\section{Relevant facts and notations} 
\label{sec:review}

\subsection{Cyclic and quasi-cyclic codes}
\label{sec:cyclic}

An $[n,k,d]_q$ code $\cal C$ linear over a finite (Galois) field
$\fq$, with $q$ a power of a prime, is a $k$-dimensional subspace of
$\fq^{n}$, the linear space of all $q$-ary strings of length $n$.  The
distance $d$ is the minimum Hamming weight of a non-zero vector in the
code, or infinity for a trivial $k=0$ code which only contains the
zero vector.  A code ${\cal C}_G\equiv {\cal C}_H^\perp$ can be
specified in terms of a generating matrix $G$ whose rows form a basis
of the code, or a parity check matrix $H$ whose rows generate the
space orthogonal to the code.

A cyclic code satisfies the additional condition that for every
codeword $c\equiv (c_{0}, c_{1}, \ldots, c_{n - 1}) \in \cal C$, its
cyclic shift $T_nc\equiv (c_{n - 1}, c_{0}, \ldots, c_{n - 2})$ also
gives a codeword, $T_nc \in \cal C$.  Shuch a shift is conveniently
represented as multiplication in the quotient polynomial ring
${\cal R}\equiv {\cal R}_{n,q} = \fq[x]/(x^n -1)$, namely,
$T_nc(x)=x c(x)\bmod x^n-1$, where
$c(x)=c_0+c_1x+\ldots+ c_{n-1}x^{n-1}$ has coefficients in $\fq^n$.  A
cyclic code is an ideal of ${\cal R}$.  In particular, this implies
that any cyclic code can be generated as the set of all multiples in
${\cal R}$ of the canonical \emph{generator polynomial} $g(x)$, where
$g(x)$ is a factor of $x^n-1$, and any such factor generates a cyclic
code.  

Both a generator and a parity check matrix (with some redundant rows)
of a cyclic code can be written as square circulant matrices.  Algebra
of circulant $n\times n$ matrices with coefficients in $\fq$ is 
isomorphic to that of polynomials in ${\cal R}$.  Indeed, given a
polynomial $a(x)\in {\cal R}$, the corresponding circulant matrix
\begin{equation} \label{eq:Cyclic-generator-matrix} 
A =  \begin{pmatrix} 
a_{0} & a_{n-1} & \ldots & a_{1}\\
a_{1} & a_{0} & \ldots & a_{2}\\
\vdots & \ddots &\ddots & \ddots\\ 
a_n& \ldots & a_1 & a_0 
\end{pmatrix}, 
\end{equation}
is conveniently written as the polynomial $A\equiv a(P)$ of the matrix
$P\equiv P_n$, the $n\times n$ cyclic permutation matrix
\begin{equation} \label{eq:permutation-matrix}
  P = \begin{pmatrix}
    0&\ldots&0 &1\\
    1&&&0\\[-0.5em]
    &\ddots &&\vdots\\
    && 1&0      
  \end{pmatrix}.
\end{equation}
We will consider vectors in $\fq^n$ as columns, so that the product
$A b$ of a circulant matrix $A=a(P)$ and a vector $b$ with the same
coefficients as in the polynomial $b(x)\in{\cal R}$ corresponds to the
product $a(x)b(x) \bmod x^n-1$.  In particular, given a canonical
generating polynomial $g(x)$, the corresponding check polynomial is
$h(x)=(x^n-1)/g(x)$, and the cyclic code
generated by $g(x)$ can
be written as%
\begin{equation}
  \label{eq:cyclic-code}
  {\cal C}_{g(x)}=\left\{c(x)\in{\cal R}: h(x) c(x)=0\bmod x^n-1\right\}.  
\end{equation}

An index-$m$ quasi-cyclic (QC) code of length $n=m\ell$ is usually
defined as a linear code invariant under the $m$-step shift
permutation $T_n^m$.  Rearranging the positions, we consider the
defining permutation as $T_\ell$ applied in each of $m$ consecutive
blocks.  As a result, a generator matrix of such a code can be written
as an $r\times n$ block matrix formed by $\ell\times \ell$ circulant
matrices.  Generally, such block matrices will be written as matrices
formed by the corresponding polynomials in ${\cal R}_{\ell,q}$.  The
same applies to vectors, which will be written as columns of
polynomials, with the exception of inline equations, where, e.g., a
two-block vector in an index-$2$ QC code may be written as
$[u(x), v(x)]$.

\subsection{Quantum CSS codes} 
\label{sec:stabilizer}

A quantum
Calderbank-Shor-Steane\cite{Calderbank-Shor-1996,Steane-1996} (CSS)
code ${\cal Q}$ with parameters $[[n,k,d]]_q$ over a Galois field
$\fq$ is isomorphic to a direct sum of an $X$- and a $Z$-like codes,
\begin{equation}
  \label{eq:css-code}
  \css(H_X,H_Z)={\cal Q}_X\oplus {\cal Q}_Z={\cal C}_{H_Z}^\perp/{\cal C}_{H_X}\oplus
  {\cal C}_{H_X}^\perp/{\cal C}_{H_Z}, 
\end{equation}
where each term in the right-hand side (r.h.s.) is a quotient of two
linear spaces in $\fq^n$, and rows of the matrices $H_X$ and $H_Z$
must be orthogonal,
\begin{equation}\label{eq:CSS-orthogonality}
  H_XH_Z^T=0.
\end{equation}
Explicitly, e.g., elements of ${\cal Q}_X$ are equivalence classes of
vectors orthogonal to the rows of the matrix $H_Z$, with any two
vectors whose difference is a linear combination of the rows of $H_X$
identified.  Vectors in the same class are called mutually degenerate,
while vectors in the class of the zero vector are called trivial.  The
codes ${\cal Q}_X$ and ${\cal Q}_Z$ have $q^k$ degeneracy classes
each, where
\begin{equation}
  \label{eq:k-CSS}
  k=n-\rank H_X-\rank H_Z
\end{equation}
is the quantum code dimension.  The distance of the code is 
$d\equiv \min(d_X,d_Z)$, where the two CSS distances, 
\begin{equation}
  \label{eq:d-CSS}
  d_X=\min_{c\in {\cal C}_{H_Z}^\perp\setminus {\cal C}_{H_X}}\wgt c,\quad 
  d_Z=\min_{c\in {\cal C}_{H_X}^\perp\setminus {\cal C}_{H_Z}}\wgt c,  
\end{equation}
are the minimum weights of non-trivial vectors (any representative) in
${\cal C}_{H_Z}^\perp$ and ${\cal C}_{H_X}^\perp$, respectively.

Physically, a quantum code operates in a Hilbert space
${\cal H}_q^{\otimes n}$ associated with $n$ quantum-mechanical
systems, \emph{qudits}, with $q$ states each, and a well defined basis
of $X$ and $Z$ operators acting in
${\cal H}_q^{\otimes
  n}$\cite{Ketkar-Klappenecker-Kumar-Sarvepalli-2006}.  Elements of
the codes ${\cal C}_{H_X}$ and ${\cal C}_{H_Z}$ correspond to $X$- and
$Z$- operators in the stabilizer group whose generators must be
measured frequently during the operation of the code; generating
matrices $H_X$ and $H_Z$ with smaller row weights result in codes
which are easier to implement in practice.  Orthogonality condition
(\ref{eq:CSS-orthogonality}) ensures that the stabilizer group is
abelian.  Non-trivial vectors in ${\cal Q}_X$ and ${\cal Q}_Z$
correspond to $X$ and $Z$ logical operators, respectively.  Codes with
larger distances have logical operators which involve more qudits;
such codes typically give better protection.

\section{Generalized Bicycle Codes}
\label{sec:GB-codes}

\subsection{Definition and general properties}
\label{sec:GB-def}

Generalized bicycle (GB)
code\cite{Kovalev-Pryadko-Hyperbicycle-2013,Panteleev-Kalachev-2019}
is a version of the bicycle
ansatz\cite{MacKay-Mitchison-McFadden-2004}, a quantum CSS code
constructed from a pair of equivalent index-two quasi-cyclic linear
codes.  Namely, given any pair of polynomials $a(x), b(x)\in F[x]$
with coefficients in a finite field $F\equiv \fq$ and of degrees
smaller than $\ell$, the generalized bicycle code $\gb(a,b)$ of
length $n=2\ell$ has CSS generator matrices specified in the block
form,
\begin{equation} \label{eq:HxHz}
  H_{X}=\left( \begin{array}[c]{c|c}A&B\end{array} \right),\quad
  H_{Z}^T=\left(\begin{array}[c]{r}B\\-A\end{array}\right).
\end{equation}
Here $A=a(P)$ and $B=b(P)$ are $q$-ary $\ell \times \ell$ circulant
matrices.  Circulant matrices necessarily commute, which guarantees
the CSS orthogonality condition (\ref{eq:CSS-orthogonality}).  For
notational convenience, we will use $[u(x), v(x)]$ to represent a
$Z$-codeword $c$, a column vector whose components in the two blocks
coincide with the coefficients of the two polynomials.  The
corresponding equation $H_Xc=0$ is equivalent to
$a(x) u(x)+b(x) v(x)=0\bmod x^\ell-1$.

With any code $\gb(a,b)$, there is an associated $q$-ary cyclic code
${\cal C}_{h(x)}^\perp\equiv \mathcal{C}_{g(x)}$ of length $\ell$,
with the check and generating polynomials 
\begin{equation}
  \label{eq:h-polynomial}
  h(x)\equiv \gcd\biglb(a(x),b(x),x^\ell-1\bigrb)\;\,\text{and}\;\,
  g(x)\equiv {x^\ell-1\over h(x)},
\end{equation}
respectively.  The
number of qudits encoded in such a GB code 
is\cite{Panteleev-Kalachev-2019}
\begin{equation}
  \label{eq:code-size}
  k=2\deg h(x),
\end{equation}
twice the dimension of the code
${\cal C}_{h(x)}^\perp\equiv {\cal C}_{g(x)}$.\\

It is easy to see that column and row permutations can be used to
obtain the matrix $H_Z$ from $H_X$, up to a sign of some columns.
Thus, the CSS distances (\ref{eq:d-CSS}) of any GB code are equal
to each other and, respectively, to the code distance $d$.  The
calculation of the distance is simplified somewhat with the help of an
auxiliary \emph{asymmetric bicycle} (AB) code
${\cal Q}'\equiv\css(H_X',H_Z)$ where
\begin{equation}
  \label{eq:AB-code-matrices}
  H_X'=\left(\begin{array}[c]{c|c}
      A_1& B_1
    \end{array}\right),\quad A_1\equiv a_1(P),\quad B_1\equiv b_1(P),
\end{equation}
where $a_1(x)\equiv a(x)/\gcd(a,b)$, $b_1(x)\equiv b(x)/\gcd(a,b)$ are
obtained by dividing the two polynomials by the common factor, and the
matrix $H_Z$ is the same as in the original GB code, see
Eq.~(\ref{eq:HxHz}).  The AB code encodes half as many qudits as the
original GB code, $k'=\deg h(x)$.  The relation between the two codes
follows from an explicit expression for the $Z$-codewords in the
original code,
\begin{equation}
  \label{eq:generic-enumerate}
 {u(x)\choose v(x)}
  =\alpha(x) g(x) {r_1(x)\choose s_1(x)}+\beta(x) {b_1(x)\choose -a_1(x)}
  \bmod x^\ell-1, 
\end{equation}
where $r_1(x)$ and $s_1(x)$ are B\'ezout coefficients such that
$a_1(x) r_1(x)+b_1(x) s_1(x)=1$ whose existence follows from
$\gcd(a_1,b_1)=1$, and, for a non-trivial codeword, at least one of
$\alpha(x)$ and $\beta(x)$ should not be divisible by $h(x)$.  Taken
separately, these two conditions yield the sets of $X$- and
$Z$-codewords of the AB code, respectively.  This results in the
following Statement whose formal proof is given in
Sec.~\ref{sec:proof-AB-code-d-eq}.
\begin{statement}
  \label{th:AB-code-d-eq}
  The distance of the code $\gb(a,b)$ is the same as that of the
  associated {\rm AB} code $\css(H_X',H_Z)$, $d=d'=\min(d_X',d_Z')$.
\end{statement}
In addition, the CSS distance $d_Z'$ (and thus the distance $d$ of the
GB code) is bounded by the distance $d_g$ of the linear cyclic code
${\cal C}_{g(x)}$.
\begin{statement}
  \label{th:AB-code-dZ-upper-bnd}
  Let $d_g$ denote the distance of the $\fq$-linear cyclic
  code with the generating polynomial $g(x)$, see
  Eq.~(\ref{eq:h-polynomial}).  Then the $Z$-distance of the $q$-ary
  AB code $\css(H_X',H_Z)$ satisfies $d_Z'\le d_g$.
\end{statement}
The formal proof in Sec.~\ref{sec:AB-code-dZ-upper-bnd} amounts to a
demonstration that for any non-zero code word
$e(x)\in {\cal C}_{g(x)}$, either $[e(x), 0]$ or $[0, e(x)]$ is a
non-trivial $Z$-vector in the AB code.

We end this section with a short list of polynomial transformations
which generate equivalent GB codes: 
\begin{statement}
  \label{th:GB-code-equivalence}
  Two codes $\gb(a,b)$ and $\gb(a',b')$ of
  the same size $n=2\ell$ are equivalent if
  \begin{enumerate}[label={\rm({\bf\roman*})}]
  \item $a'(x)=a(x^m)\bmod x^\ell-1$, $b'(x)=b(x^m)\bmod x^\ell-1$ for
    some $m$ mutually prime with $\ell$, $\gcd(m,\ell)=1$;
  \item $a'(x)=b(x)$, $b'(x)=a(x)$; 
  \item $a'(x)$ and $b'(x)$ are the reciprocal polynomials of $a(x)$
    and $b(x)$, respectively.
  \item $a'(x)=\delta a(x)$, $b'(x)=b(x)$, for some $0\neq\delta\in\fq$.
  \item $a'(x)=f(x) a(x)$, $b'(x)=f(x) b(x)$, for some polynomial
    $f(x)\in\fq[x]$ such that $\gcd(f,x^\ell-1)=1$.
  \end{enumerate}  
\end{statement}
The first four transformations correspond to permutations preserving
the circulant symmetry\cite{MS-book}, while the last one may be useful
for constructing LDPC codes, since minimum row weight does not
necessarily correspond to minimum polynomial degrees.

While technically not an equivalence transformation, we should also
mention here the case of polynomials \emph{commensurate} with the
circulant size $\ell$, i.e., such that $h(x)=h_0(x^\Delta)$, where
$\Delta>1$ is a factor of $\ell$.  A cyclic code whose check
polynomial $h(x)$ is commensurate with $\ell$ is merely a direct sum
of $\Delta$ disconnected cyclic codes, each equivalent to the code of
length $\ell_0\equiv \ell/\Delta$ with the check polynomial $h_0(x)$.
Same is true in the case of a code $\gb(a,b)$ whose defining
polynomials have the same commensurability factor $\Delta$:
\begin{statement}[Commensurate GB code]
  A code $\gb(a,b)$ with parameters $[[2\ell,k,d]]_q$ and 
  $a(x)=a_0(x^\Delta)$, $b(x)=b_0(x^\Delta)$, where
  $\ell=\ell_0\Delta$, is equivalent to a direct sum of $\Delta$
  copies of the code $\gb(a_0,b_0)$ with parameters
  $[[2\ell_0,k_0,d_0]]_q$.  In particular, $d=d_0$ and $k=k_0\Delta$.
\end{statement}
A cyclic or GB code that is not commensurate is called
\emph{incommensurate}.

\subsection{Bounds for GB codes of unrestricted weight}
\label{sec:GV-bounds}

Here we give several existence bounds for general (non-LDPC) GB codes,
using the standard map\cite{Calderbank-Shor-1996,Steane-1996,%
  Ketkar-Klappenecker-Kumar-Sarvepalli-2006} relating the parameters
of a CSS code to those of the associated pair of classical
$\fq$-linear mutually dual-containing codes.  In the case of the code
$\gb(a,b)$, the two codes have double-circulant parity check matrices
$H_X$ and $H_Z$ given in Eq.~(\ref{eq:HxHz}).  To be specific, we
focus on the index-two QC code with the check matrix $H=H_X$, and
denote such a code QC$(a,b)$.
\begin{statement}[CSS map for GB codes\cite{Calderbank-Shor-1996,%
    Steane-1996,Ketkar-Klappenecker-Kumar-Sarvepalli-2006}]
  \label{th:qc-map}
  Given the parameters $[n_0=2\ell,k_0,d_0]_q$ of the classical linear
  code {\rm QC}$(a,b)$, the quantum CSS code $\gb(a,b)$ has parameters
  $[[2\ell,2k_0-2\ell,d]]_q$, where $d\ge d_0$.
\end{statement}
It is a classical result\cite{Chen-Peterson-Weldon-1969,Kasami-1974}
that index-two QC codes include good codes with rate $1/2$ and
asymptotically finite relative distances $d_0/n_0\to \delta_0>0$.
However, the codes used in the proof have parity-check matrices in a
systematic form with $A=I$; for such a self-dual (up to a permutation)
index-two QC code Statement \ref{th:qc-map} gives a quantum code which
encodes no qudits.  A number of other lower bounds on the distances of
QC codes have been constructed, in particular, a
version\cite{Semenov-Trifonov-2012} of the BCH bound (for a recent
review, see Ref.~\onlinecite{Guneri-Ling-Ozkaya-2020}).  However, none
of these bounds gives a family of QC codes with
$k_0-\ell=\mathcal{O}(\ell)$ and $d_0=\mathcal{O}(n)$.  Indeed, by
Statements \ref{th:AB-code-d-eq} and \ref{th:AB-code-dZ-upper-bnd},
such a family of QC codes would imply that linear cyclic codes must be
asymptotically good, a question which remains
unresolved\cite{Lin-Weldon-1967,MartinezPerez-Willems-2006}.

For these reasons here we list several partial results, which
demonstrate the existence of QC codes with sublinear $k_0-\ell$ and
distances scaling linearly, and of finite-rate QC codes with sublinear
(power law) distances.  The following bound is constructed using
elementary arguments similar to those used in
Ref.~\onlinecite{Galindo-Hernando-Matsumoto-2018}:
\begin{statement}
 \label{th:bound-qc-special}
 Consider the code {\rm QC}$(a,b)$ in the special case
 $a(x)=f(x)h(x)$, $b(x)=h(x)$, where for some polynomial $r(x)$,
 $\gcd\biglb(f(x)-r(x),x^\ell-1\bigrb)=p(x)$ is a factor of the
 generating polynomial, $g(x)=p(x)q(x)$.  Then the distance of the QC
 code satisfies the bounds:
  \begin{enumerate}[label={\rm(\alph*)}]
  \item If $r(x)=0$, \quad
    $d_0 \ge\min\bigl\{ d[q],1+d[p]\bigr\}$;
  \item Otherwise, if $\gcd\biglb(r(x),x^\ell-1\bigrb)=1$,
    $$d_0 \ge\min\Bigl\{2d[q ],
      d[p]/\wgt(r)\Bigr\}.$$
\end{enumerate}
Here $h(x)$ and $g(x)$ are given by Eq.~(\ref{eq:h-polynomial}), and
$d\left[q\right]$ is the distance of the linear cyclic code
generated by $q(x)$.
\end{statement}
Unfortunately, the codes generated by $p(x)$ and $q(x)=g(x)/p(x)$,
respectively, form a pair of dual-containing cyclic codes; it is well
known\cite{Aly-Klappenecker-Sarvepalli-2007} that the minimum of the
two distances is bounded by $\mathcal{O}(\sqrt\ell)$, which limits the
usability of the bound in Statement \ref{th:bound-qc-special}.

The following bound obtained with the help of a counting argument is a
variant of Lemma 5 from Ref.~\onlinecite{Kovalev-Dumer-Pryadko-2011}
in application to GB codes:
\begin{statement}\label{th:gv-like-bnd}
  Let $x^\ell -1=g(x)h(x)$ with $g(x)\in\fq[x]$ irreducible, and
  \begin{equation}
    \label{eq:gv-sum}
    d_{\rm GV}=\max d:    \sum_{s=1}^{d-1}(q-1)^s\left[{2\ell\choose s}
      -{\ell\choose s}\right]<q^{\ell-\deg h}-1. 
  \end{equation}
  Then, there exists $f(x)\in\fq[x]$ such that the length-$2\ell$ code {\rm
    QC}$(hf,h)$ has distance
  $d\ge \min(d[g],d_{\rm GV})$, where $d[g]$ is the distance of the
  cyclic code generated by $g(x)$.
\end{statement}
The counting part of this bound asymptotically approaches from above
the Gilbert-Varshamov (GV) bound\cite{Gilbert-1952,Varshamov-1957} for
linear $q$-ary codes with $k=\ell+\deg h$, which coincides with the GV
bound\cite{Calderbank-Shor-1996} for $q$-ary CSS codes with
$k=2\deg h$.  Unfortunately, the requirement for $g(x)$ to be
irreducible is very restrictive.  Generally, since $x^{ab}-1$ has both
$x^a-1$ and $x^b-1$ as factors, codes with $\ell$ prime get higher
lower bounds on their relative distances under Statement
\ref{th:gv-like-bnd}.  In particular, two well-known special cases
correspond to $x^\ell-1$ having only two and three factors,
respectively:
\begin{example}
  \label{th:linear-distance-QC-codes}{\rm [GB codes with linear distance]}
  Let $\ell$ be such that $\ord_\ell(q)=\ell-1$, where $\ord_\ell(q)$
  is the multiplicative order function of $q$ modulo $\ell$.  This
  ensures that $x^\ell-1$ has only two irreducible factors in
  $\fq[x]$, $h(x)\equiv 1-x$ and $g(x)=1+x+\ldots +x^{\ell-1}$.  Then
  there is a {\rm GB} code with parameters
  $[[2\ell,2,d\ge d_{\rm GV}]]_q$, where $d_{\rm GV}$ is given by
  Eq.~(\ref{eq:gv-sum}).  For $q=2$ the corresponding set
  is\cite{OEIS-seq-A001122}
  $\{3, 5, 11, 13, 19, 29, 37, 53, 59, 61, 67, 83, 101, 107, 131,
  139$, $149, 163, 173, 179, 181, 197, \ldots\}$, and, moreover,
  according to Artin's primitive root conjecture, a finite fraction of
  all primes satisfies this condition for any $q>0$ which is not a
  perfect square\cite{HeathBrown-1986}.  Asymptotically, at
  $\ell\to\infty$, this bound on the relative distance coincides with
  the {\rm GV} bound for rate-$1/2$ linear $q$-ary codes, e.g.,
  $\delta_{\rm GV}\approx 0.1100$ for $q=2$.
\end{example}
\begin{example}
  \label{th:quadratic-residue-QC-codes}{\rm [GB codes with asymptotic rate
  $1/4$]}
  For an odd prime $\ell$ let a prime $p$ be a quadratic residue
  modulo $\ell$, i.e., $p\equiv m^2\bmod \ell$ for some integer $m$.
  Then, $x^\ell-1$ has only three irreducible factors in $\fp[x]$, and
  there is a \emph{quadratic-residue} cyclic code
  $[\ell,(\ell+1)/2,d]_p$ with $d\ge \sqrt{\ell}$ and an irreducible
  generator polynomial\cite{MS-book}. According to Statement
  \ref{th:gv-like-bnd}, a prime-field {\rm GB} code with parameters
  $[[2\ell,(\ell-1)/2,d\ge \ell^{1/2}]]_p$ exists.
\end{example}

\subsection{A map to hypergraph-product and related codes}
\label{sec:QHP-map}

\begin{figure}[htbp]
  \centering
({\bf a}) \includegraphics[width=0.7\columnwidth]{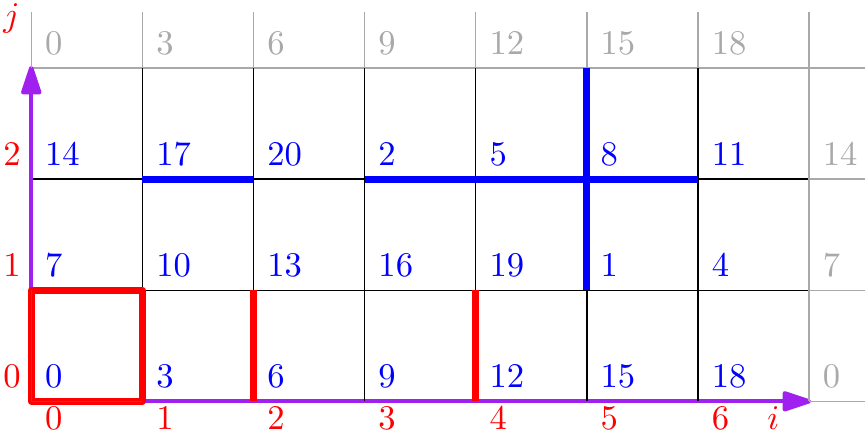}\hfill\\
({\bf b}) \includegraphics[width=0.7\columnwidth]{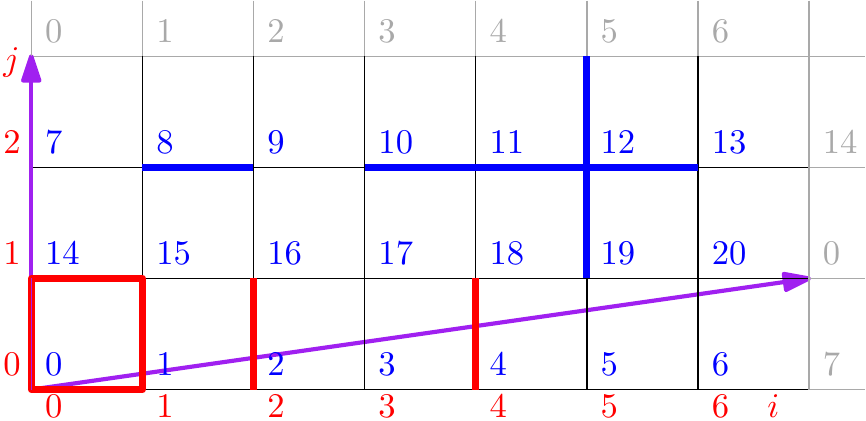}
\caption{(Color online) ({\bf a}) A map (\ref{eq:qhp-map}) of an
  $n_1\times n_2$ square lattice with periodic boundary conditions
  along the vectors $\vec L_1=(n_1,0)$ and $\vec L_2=(0,n_2)$ to a
  chain of length $\ell=n_1n_2$, with $n_1=7$ and $n_2=3$. Red digits
  below and to the left of the axes show the column $i$ and row $j$
  indices; the index $t$ is placed above and to the right of the
  corresponding vertex.  The two blocks in Eq.~(\ref{eq:hp-two-circ})
  correspond to horizontal and vertical edges, respectively.  Thicker
  red and blue edges, respectively, indicate those in an $X$ and a $Z$
  stabilizer generators of the QHP code $[[42,8,3]]$ obtained from
  polynomials $h_1(x)=1+x+x^2+x^4$ and $h_2(x)=1+x$. The equivalent
  code $\gb(a,b)$ has $a(x)=1+x^3+x^6+x^{12}$ and $b(x)=1+x^7$.  ({\bf
    b}) Same, but with a skewed periodicity vector
  $\vec L_1'=(n_1,1)$.  The corresponding map $t=i-n_1j\bmod n_1n_2$
  is invertible, but has a different symmetry.  As a result, even
  though the GB code with $a'(x)=1+x+x^2+x^4$ and $b'(x)=1+x^{14}$ has
  the same parameters $[[42,8,3]]$, this is coincidental.  Indeed,
  replacing the polynomial $h_2(x)$ with $h_2'(x)=1+x+x^2$ gives the
  QHP code $[[42,16,2]]$ and an equivalent code using the map
  (\ref{eq:qhp-map}), but the present map gives $b''(x)=h_2'(x^7)$
  which is mutually prime with $a(x)$, resulting in an empty GB code.}
\label{fig:hp-gb-map}
\end{figure}

We would now like to focus on more practical GB codes with
bounded-weight stabilizer generators.  First, we construct an explicit
map between a quantum hypergraph-product code\cite{Tillich-Zemor-2009}
constructed from a pair of square circulant matrices of mutually prime
dimensions $n_1$ and $n_1$, and a GB code with circulant size
$\ell=n_1n_2$, see Fig.~\ref{fig:hp-gb-map}.

\begin{figure}[htbp]
  \centering
  \includegraphics[width=0.97\columnwidth]{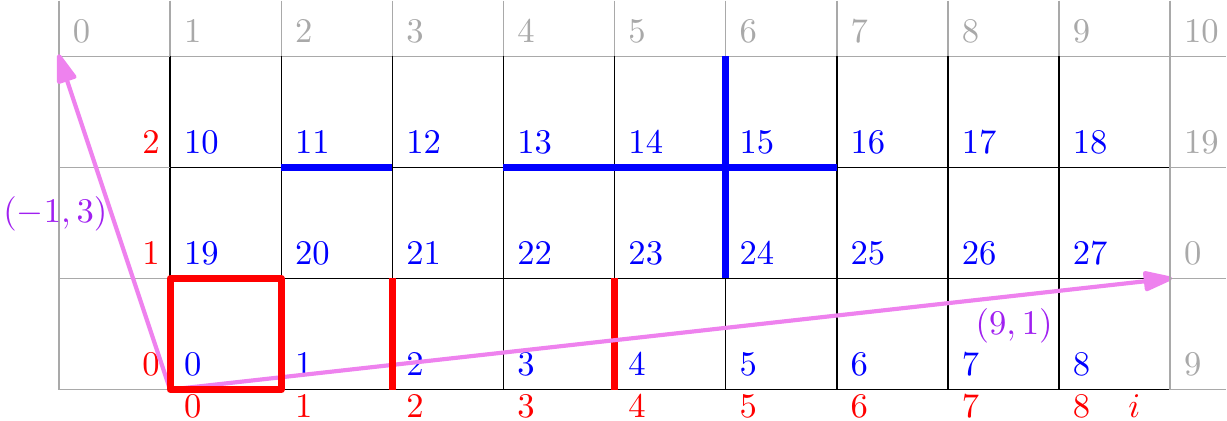}
  \caption{(Color online) As in Fig.~\ref{fig:hp-gb-map} but with
    periodicity vectors $\vec L_1=(9,1)$, $\vec L_2=(-1,3)$ and 
    circulant matrices of size $\ell=|\vec L_1\times \vec L_2|=28$.
    Here stabilizer generators with the lattice structure identical to
    those in Fig.~\ref{fig:hp-gb-map} give a rotated-QHP code
    $[[56,2,8]]$.  The equivalent code $\gb(a,b)$ has
    $a(x)=1+x+x^2+x^{4}$ and $b(x)=1+x^{19}$.}
\label{fig:rhp-gb-map}
\end{figure}

Specifically, let $H_1=h_1(P_{n_1})$ and $H_2\equiv h_2(P_{n_2})$ be a
pair of square circulant matrices of size $n_1$ and $n_2$,
corresponding to polynomials $h_1(x)$ and $h_2(x)$ in $\fq[x]$,
respectively.  Given the parameters $[n_i,k_i,d_i]_q$ for the two
cyclic codes with the check polynomials $h_i$, $i\in\{1,2\}$, consider
the hypergraph-product code with CSS generators in a block form
written as Kronecker products,
\begin{eqnarray}
  \label{eq:hp-two-circ}
  H_X=(I_1\otimes H_2, H_1\otimes I_2),\;\; H_Z^T=\left(
  \begin{array}[c]{c}
    H_1\otimes I_2\\ -I_1\otimes H_2,
  \end{array}\right),\quad 
\end{eqnarray}
where $I_i$ are the identity matrices of size $n_i$, $i\in \{1,2\}$.
Such a code has the
parameters\cite{Tillich-Zemor-2009,Zeng-Pryadko-hprod-2020}
\begin{equation}
[[2n_1n_2,2k_1k_2,\min(d_1,d_2)]]_q\label{eq:hp-two-circ-params}
\end{equation}
and can be put on an $n_1\times n_2$ square lattice with periodic
boundary conditions as illustrated in Fig.~\ref{fig:hp-gb-map}(a),
with the two blocks in Eq.~(\ref{eq:hp-two-circ}) corresponding to
qubits on horizontal and vertical edges, respectively.

In the special case where $n_1$ and $n_2$ are mutually prime,
$\gcd(n_1,n_2)=1$, an equivalent GB code with circulant size
$\ell=n_1n_2$, can be constructed from the polynomials
\begin{equation}
  \label{eq:GB-HP-code}
a(x)=h_1(x^{n_2}),\quad b(x)=h_2(x^{n_1}), 
\end{equation}
where the values of the circulant index
\begin{equation}
t=n_2 i+n_1 j\bmod \ell\label{eq:qhp-map}
\end{equation}
are in a one-to-one correspondence with the positions $(i,j)$ on the
$n_1\times n_2$ portion of the square lattice with periodic boundary
conditions introduced by identifying any pair of points connected by
periodicity vectors $\vec L_1=(n_1,0)$ and $\vec L_2=(0,n_2)$.

We should emphasize that in addition to being a one-to-one map,
Eq.~(\ref{eq:qhp-map}) has the correct translation symmetry.
Different GB codes can be also obtained using skewed periodicity
vectors, e.g., $\vec L_1'=(n_1,1)$ instead of $\vec L_1$, equivalent
to the index map $t=i-n_1j\bmod \ell$.  This map does not give
identity transformation for the translation $i\to i+n_1$.  Thus, we do
not expect the corresponding code $\gb(a',b')$, $a'(x)=h_1(x)$,
$b'(x)=h_2(x^{n_1})$ to be equivalent to the original QHP code, see
Fig.~\ref{fig:hp-gb-map}({\bf b}).

Generally, a quantum code on the edges of a square lattice with
stabilizer generators similar to those of a QHP code but with
periodicity vectors non-collinear with the axes is called a rotated
QHP code\cite{Kovalev-Pryadko-Hyperbicycle-2013}, a code in a more
general class of lifted-product codes\cite{Panteleev-Kalachev-2020}.
\begin{statement}
  \label{th:rotated-QHP-map}
  An arbitrary GB code of length $2\ell$ is equivalent to a rotated
  QHP code with periodicity vectors $\vec L_1$ and $\vec L_2$ such
  that $|\vec L_1\times \vec L_2|=\ell$.
\end{statement}
\begin{proof}
Indeed, given a decomposition $\ell=n_1n_2+\lambda$, where $n_1$ and
$\ell$ are mutually prime, $\gcd(n_1,\ell)=1$, consider a pair of
vectors,
\begin{equation}
  \vec L_1=(n_1,1) \;\text{ and }\;
  \vec L_2=(\lambda,n_2).\label{eq:periodicity-vectors}
\end{equation}
If we use these as periodicity vectors (i.e., identify any pair of
points on the square lattice connected by one of these vectors), there
are exactly $\ell=|\vec L_1\times \vec L_2|$ inequivalent points with
a one-to-one map $t=i-n_1j\bmod \ell$ to a cycle $\mathbb{Z}_\ell$,
see Figs.~\ref{fig:hp-gb-map}({\bf b}) and \ref{fig:rhp-gb-map}.
Then, given the polynomials $h_1(x)$ and $h_2(x)$ which define the
lattice layout of the stabilizer generators of a rotated QHP code with
the chosen periodicity vectors, the polynomials defining the
corresponding GB code are $a(x)=h_1(x)$ and $b(x)=h_2(x^{n_1})$.

Conversely, let $m_1$ be a multiplicative inverse of $n_1$ modulo
$\ell$, $m_1n_1=1\bmod \ell$; its existence is guaranteed by the
condition $\gcd(n_1,\ell)=1$.  Then, given the code $\gb(a,b)$, we
recover the polynomials for the corresponding rotated-QHP code,
$h_1(x)=a(x)$ and $h_2(x)=b(x^{m_1})\bmod x^\ell-1$.
\end{proof}
These maps show, in particular, that GB codes can be as good as QHP
codes constructed from two square circulant matrices of mutually prime
sizes.  Given the explicit Eq.~(\ref{eq:hp-two-circ-params}) relating
parameters of a QHP code with those of the two cyclic codes with
parity-check polynomials $h_1(x)$ and $h_2(x)$, we obtain an existence
for GB codes of finite rates and a power-law distance scaling as
$\mathcal{O}\biglb(n^{1/2}/\polylog(n)\bigrb)$ or better.  Indeed, the
question of whether long linear cyclic codes are asymptotically good
is still open, with only minor progress made in recent
years\cite{MartinezPerez-Willems-2006,Haviv-Langberg-Schwartz-Yaakobi-2017,%
  Shi-Wu-Sole-2018}.  In reality, the question is academic, since
finite-length performance of cyclic codes is excellent, and already
the BCH bound gives codes\cite{Berlekamp-1972} with rate $R>0$ and
$\delta\ge (2\ln R^{-1})/\log n$, while linear cyclic codes with
$\delta>(1-2R)/\sqrt{2\log n}$ can also be
constructed\cite{Berlekamp-Justesen-1974}.

From a practical viewpoint, more interesting are the bounds on
parameters of LDPC GB codes with stabilizer generators of bounded
weight.  We construct such (upper) bounds in the next section with the
help of general results by Bravyi, Poulin, and
Terhal\cite{Bravyi-Terhal-2009,Bravyi-Poulin-Terhal-2010}, by mapping
a linear cyclic code with check polynomial of weight $w_1$ to a code
local on a $D$-dimensional hyper-cubic lattice, with $D\le w_1$, and a
GB code with row weight $w$ to a quantum code local on a
$D$-dimensional lattice, with $D\le w-1$.

\subsection{A map to a code local in $D$ dimensions}
\label{sec:D-map}

Let us first consider the case of a cyclic code of length $\ell$ with
the parity check polynomial $h(x)\in\fq[x]$ of a fixed weight $w$.
Here we will not require that $h(x)$ be a factor of $x^\ell-1$, as
such factors do not necessarily have minimal weights, but a $q$-ary
polynomial such that the canonical check polynomial
$h_1(x)\equiv\gcd(h,x^\ell-1)$ be non-trivial, $k=\deg h_1(x)>0$.

The following is a generalization of Statement
\ref{th:rotated-QHP-map}:
\begin{statement}
  \label{th:cyclic-code-map}
  An incommensurate linear cyclic code of length $\ell$ with check
  polynomial $h(x)$ of weight $w$ is equivalent to a code with all
  checks local on a hypercubic lattice of dimension $D\le w$, and
  $D\le w-1$ if $\ell$ is prime.
\end{statement}
\begin{proof}
  For a polynomial $h(x)$ with monomial degrees
  $0=t_0<t_1<\ldots < t_{w-1}$, consider a set of $w$ integer vectors
  in $\mathbb{Z}^{ w}$, written as the rows of the
  lower-triangular matrix
  \begin{equation}
    M=\left(
      \begin{array}[c]{ccccc}
        \ell  &  & &\\
        t_1   &-1& &\\
        t_2   &  &-1&\\
        \vdots&  & &\ddots\\
        t_{w-1}&  & & &-1
      \end{array}
    \right).
    \label{eq:transf-matrix}  
  \end{equation}
  The determinant of $M$ equals $\pm\ell$, and by the
  incommensurability condition, there exists a map from the chain
  $0\le t<\ell$ to the region in $\mathbb{Z}^{w}$ given by the
  inequalities $0\le x_i<\ell_i$, $0\le i<w$, where $\ell_0=t_1$,
  $\ell_i=\lceil t_{i+1}/t_i\rceil$ for $0<i<w-1$, and
  $\ell_{w-1}=\lceil\ell/t_{w-1}\rceil$.  With these notations, the
  check polynomial becomes $a_0+a_{t_1} x_1+\ldots a_{t_{w-1}}x_{w-1}$, i.e.,
  the checks are one-local in the bulk of the region (with the
  structure as in quantum fractal
  codes\cite{Yoshida-topo-2013,Kalachev-Panteleev-2020}), and at most
  two-local near the region's boundary.

  When $\ell$ is a prime (or one of the original degrees $t_i\neq0$ is
  mutually prime with $\ell$), there exists $m\in\mathbb{Z}_\ell$ such
  that $mt_i=1\bmod \ell$, and $x^i\to x^{im}\bmod x^\ell-1$ gives an
  equivalent code, see Statement \ref{th:GB-code-equivalence}.  The
  modified check polynomial $h'(x)\equiv h(x^m)\bmod x^\ell-1$ $h'(x)$
  has a degree-one monomial, and the region defined by the periodicity
  vectors (\ref{eq:transf-matrix}) has $x_0=x_1$, thus $D\le w-1$.

  The dimension can be additionally reduced if there is a simple
  relation between the monomial degrees, e.g., $t_3=t_1+t_2$, in which
  case the third axis can be skipped and the corresponding monomial
  written as $a_{t_3} x_1x_2$.
\end{proof}
Given such a map to a code local in $D$ dimensions, with the help of
the general result in the appendix of
Ref.~\onlinecite{Bravyi-Poulin-Terhal-2010}, we immediately obtain:
\begin{corollary}
  \label{th:upper-d-bnd-classical}
  Parameters $[\ell, k_1,d_1]_q$ of any $\fq$-linear cyclic code of
  length $\ell$ with the check polynomial of weight $w_1$ which is
  equivalent to a code local in $D_1\le w_1$ dimensions, satisfy
  $k_1d_1^{1/D_1}=\mathcal{O}(\ell)$.
\end{corollary}

The case of a GB code with polynomials $a(x)$ and $b(x)$ with the
total weight $w$ is considered similarly, except that each vertex of
the hypercubic lattice must now contain two qudits, one from each
block, and the maximum dimension is additionally reduced by one since
both polynomials have zero-degree monomials.  It is also easy to check
that a local map for $H_X$ to $\mathbb{Z}^{D}$ automatically implies
the locality of the corresponding $H_Z$.  We have, combining the
results from
Refs.~\onlinecite{Bravyi-Terhal-2009,Bravyi-Poulin-Terhal-2010}:
\begin{statement}
  \label{th:upper-d-bnd-GB}
  An incommensurate GB code with row weight $w$ and parameters
  $[[n=2\ell, k,d]]_q$ is equivalent to a CSS code local in $D\le w-1$
  dimensions ($D\le w-2$ if $\ell$ is prime).  Its parameters
  satisfy the inequalities $$d\le \mathcal{O}(n^{1-1/D})\ \ \text{\rm and}\ \ 
  kd^{2/{(D-1)}}\le \mathcal{O}(n).$$
\end{statement}
Notice that the last equation implies that any GB code family with a
fixed weight $w$ has an asymptotically zero rate, since $k/n\to 0$
when the distance $d$ becomes infinite. 

\subsection{Exact bound for GB codes of weight four}
\label{sec:weight-four}

Here we consider in detail the special case of codes with $w=4$.
According to Statement \ref{th:upper-d-bnd-GB}, any such code is
equivalent to a code local in two dimensions.  The case of $D=2$ is
special, since
Refs.~\onlinecite{Bravyi-Terhal-2009,Bravyi-Poulin-Terhal-2010} give
asymptotically exact bounds for such codes.

A non-trivial GB code of weight $w=4$ can only be constructed when
both $a(x)$ and $b(x)$ have equal weights.  Moreover, weight-two
polynomials of equal degrees, or a polynomial of degree $\ell/2$ with
$\ell$ even, always give an empty code or a distance-two code.
Therefore, for a non-trivial incommensurate GB code with distance
$d\ge3$, with the help of Statement \ref{th:GB-code-equivalence},
without restricting generality, we can request that the degrees
$\alpha= \deg a(x)$ and $\beta=\deg b(x)$ satisfy
$\alpha<\beta<\ell/2$, with $\gcd(\alpha,\beta,\ell)=1$.

These additional properties guarantee that any pair of rows of a
generator matrix $H_X$ (or $H_Z$) in Eq.~(\ref{eq:HxHz}) intersect in
at most one column, and any column has exactly two non-zero elements,
as in a vertex-edge incidence matrix of a simple graph.  The analogy
can be made exact by considering a pair of binary matrices $J$, $F$
constructed from $H_X$ and $H_Z$, respectively, by replacing any
non-zero element with $1$.  The rows of the two matrices are
necessarily orthogonal, $JF^T=0$ (over $\mathbb{Z}_2$).  Thus, these
matrices can be readily identified as a vertex-edge and a face-edge
incidence matrices of a locally planar $(4,4)$ graph ${\cal G}$, i.e.,
with each vertex of ${\cal G}$ and the corresponding dual graph
$\widetilde{\cal G}$ of equal degree $4$.  Finally, it is also easy to
see that the graph ${\cal G}$ is locally (i.e., as long as the current
position does not close a circle $t\to t+\ell$) isomorphic to a square
lattice, with the two blocks, respectively, corresponding to
horizontal and vertical edges, and oriented in the direction of
increasing index.  Namely, any (local) sequence of horizontal
$x_i=\pm1$ and vertical $y_j=\pm1$ steps, where the signs indicate the
direction, arrives at the same final position as long as the total
displacements $\sum x_i$ and $\sum y_j$ coincide.  That is, the graph
${\cal G}$ is \emph{covered} by the infinite square lattice graph
${\cal H}$, with the covering function $f:{\cal H}\to {\cal G}$ such
that a path between a pair of vertices on ${\cal H}$ with the same
covering map image corresponds to a non-trivial cycle on ${\cal G}$,
or one or more ``large'' displacements $t\to t\pm \ell$ of the
circulant index.

With such a map, it is evident that a non-trivial GB code of
weight-four and distance $d\ge3$ is a square-lattice surface code,
with $Z$-codewords corresponding to homologically non-trivial cycles,
with the homology fixed by the covering map $f$ (see, e.g.,
Ref.~\onlinecite{Woolls-Pryadko-2020}).  Then, the distance $d_Z$ is
the length of a shortest path connecting a pair of distinct vertices
on ${\cal H}$ whose covering-map images coincide on ${\cal G}$.

\begin{figure}[htbp]
  \centering
  \includegraphics[width=0.7\columnwidth]{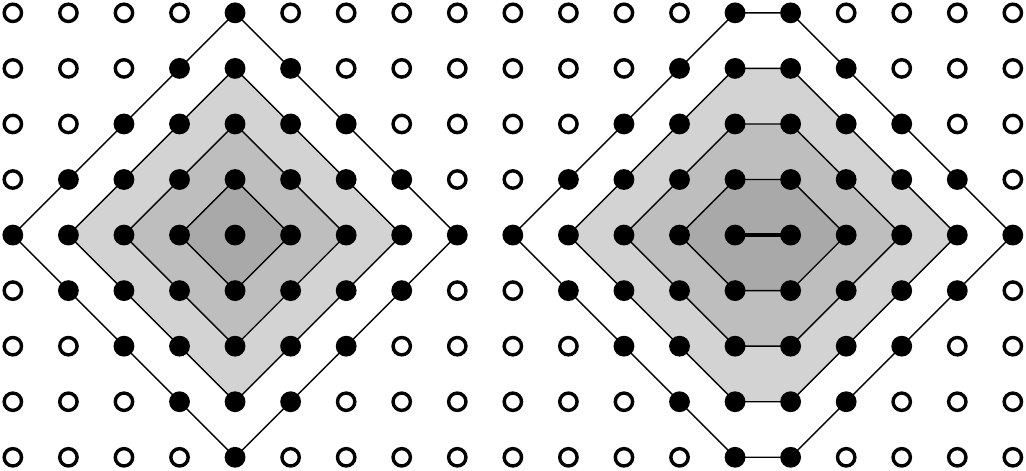}
  \caption{Left: squares with progressively ligher shading indicate
    vertex-centered balls of radius $r=1$, $2$, $3$, and $4$ on the
    square lattice; the numbers of vertices on the boundary are $4$,
    $8$, $12$, and $16$, respectively.  Right: same for edge-centered
    regions where each distance-$r$ boundary has exactly two
    additional vertices.}
  \label{fig:sq-neig}
\end{figure}

To construct an actual distance bound, start with an arbitrary vertex
$i\in{\cal V}_{\cal H}$ (where ${\cal V}_{\cal H}$ is the vertex set
of ${\cal H}$), and consider a vertex-centered ball
$B_r(i)$ on ${\cal H}$, a set of all vertices
$j\in{\cal V}_{\cal H}$ such that the graph distance $d(i,j)\le r$,
see Fig.~\ref{fig:sq-neig} (left).  With the circulant size $\ell$, the graph
${\cal G}$ has exactly $\ell$ vertices.  Thus, if the size of the ball
satisfies $|B_r(i)|>\ell$, the ball must include at least two
equivalent vertices, which gives for the code distance, $d_Z\le 2r$,
the \emph{diameter} of the ball.  The size of a ball on the square
lattice is computed easily by summing the arithmetic sequence,
$$
|B_r(i)|-1=4+8+\ldots+ 4r=2 r(r+1),
$$
which gives the upper bound $d_Z\le 2r$ for any circulant size
$\ell < 1 + 2r(r + 1)$.  A similar calculation for an edge-centered
ball on ${\cal H}$ gives an odd-valued upper bound $d_Z\le 2r+1$ for
any $\ell<2(r+1)^2$, see Fig.~\ref{fig:sq-neig} (right).  We rewrite
these inequalities equivalently as lower bounds on the code length
$n=2\ell$ for a given value of the distance $d=d_Z$:
\begin{statement}
  \label{th:sq-surface-code-bounds}
  Consider a weight-four GB code of an odd distance $d=2r+1$, then its
  length $n\ge 1+d^2$.  For an even distance $d=2r$, the length
  $n\ge d^2$.
\end{statement}
The argument above is valid for $d\ge3$.  We verified by exhaustive
search that these inequalities are also valid for $d\in\{1,2\}$. 

We notice that the inequalities in Statement
\ref{th:sq-surface-code-bounds} are sharp for surface codes.  Namely,
the odd-distance bound is reached by a
family\cite{Kovalev-Pryadko-2012} of square lattice surface codes with
periodicity vectors $(r+1,r)$ and $(-r,r+1)$ and parameters
$[[(2r+1)^2+1,2,2r+1]]_q$, while the even-distance bound is achived by
the $45^\circ$-rotated surface codes\cite{Bombin-2007}.  These latter
codes have periodicity vectors $(\pm r,\pm r)$ and parameters
$[[4r^2, 2, 2r]]_q$.  However, the corresponding translation group
$\langle x,y\mid xyx^{-1}y^{-1}=x^ry^r=x^ry^{-r}=1\rangle$ is not
cyclic for any $r>1$, which proves that there are no corresponding GB
codes except for $r=1$, with parameters $[[4,2,2]]_q$.

The next-shortest family of even-distance surface codes has
periodicity vectors $(r\pm1,1\pm r)$ and parameters
$[[4r^2+4,2,2r]]_q$, $r\in\mathbb{N}$; these have GB code
representations when $r$ is even, which requires the distance $2r$ be
a multiple of four.

\section{Numerical results}
\label{sec:numeric}

To summarize our results so far, we expect the highest distances for
GB codes encoding $k=2$ qudits, with $b(x)=1+x$ and $a(x)$ of even
weight, which ensures the corresponding check polynomial
(\ref{eq:h-polynomial}) to be $h(x)=1+x$ for any $\ell\ge2$.  For the
qubits (quantum codes over the binary field $\mathbb{Z}_2$), Example
\ref{th:linear-distance-QC-codes} based on Statement
\ref{th:gv-like-bnd} shows that for prime circulant sizes $\ell$ with
a primitive root $2$, GB codes in this family exist with relative
distance $d/n>\delta_{\rm GV}\approx 0.11$.  However, the upper and
lower bounds for the codes of row weight $w$ (which corresponds to
$\wgt(a)=w-2$) differ strongly for $w>4$.  Namely, Statement
\ref{th:bound-qc-special}, the map to QHP codes in
Sec.~\ref{sec:QHP-map}, and several explicit $w=4$ code families in
Sec.~\ref{sec:weight-four} agree that such codes with the distances
$d> \mathcal{O}( n^{1/2})$ scaling as a square root of the block size
exist.  On the other hand, the upper bound in Statement
\ref{th:upper-d-bnd-GB} for such codes suggests a power-law distance
scaling with the exponent that may change with $w$,
$d< \mathcal{O}( n^{\gamma})$, where $\gamma=1-1/D$, with the
effective dimension $D(w)\le w-2$ for a prime $\ell$.  The two bounds
give the same exponent $\gamma=1/2$ only for $w=4$, while there is an
interval of possible exponent values for $w>4$.  Notice that any
exponent, including $\gamma_{\rm min}=1/2$, may be consistent with the
linear distance scaling at large $w$, if the corresponding prefactor
$A(w)$ in the power-law $d\propto A(w)n^\gamma$ diverges at
$w\to\infty$.

To address this issue, we set up to find largest-distance GB codes
based on qubits and row weights $w\in\{4,6,8\}$, fixing $b(x)=1+x$.
Namely, for every prime $\ell\le 227$ such that $2$ is a primitive
root, we calculated the maximum distance of GB codes over inequivalent
polynomials $a(x)$ of weights $2$, $4$, and $6$ (also, for every prime
$\ell\le 127$ in the case of $\wgt a=4$, which did not substantially
modify the results).  We used equivalence maps (\textbf{iii}) and
(\textbf{v}) [with $f(x)=x^s$, $s<\ell$] in Statement
\ref{th:GB-code-equivalence} to define a canonical form of
$a(x)\in \mathbb{F}_2[x]$ of degree $\Delta$, with $a_0=a_\Delta=1$,
and smallest alphabetically.  In particular, this implies a
smallest-degree polynomial in each equivalence class.  When
enumerating polynomials, we discarded any which did not coincide with
the corresponding canonical form.  Actual distance calculation were
done using the GAP package {\tt
  QDistRnd}\cite{Pryadko-Shabashov-Kozin-QDistRnd-2021}, with the help
of the auxiliary AB code as in Statement \ref{th:AB-code-d-eq}, and
only for those polynomials $a(x)=f(x)(1+x)$ with a sufficiently large
$1+\wgt f$ (such an upper bound on the distance is a trivial
consequence of Statement \ref{th:GB-code-equivalence}).  The resulting
data and the actual codes are available for download at the GitHub
repository {\tt QEC-pages/GB-codes}\cite{Wang-Pryadko-GBcode-2022}.

\begin{figure}[htbp]
  \centering
  \includegraphics[width=\columnwidth,bb=0 5 338 205]{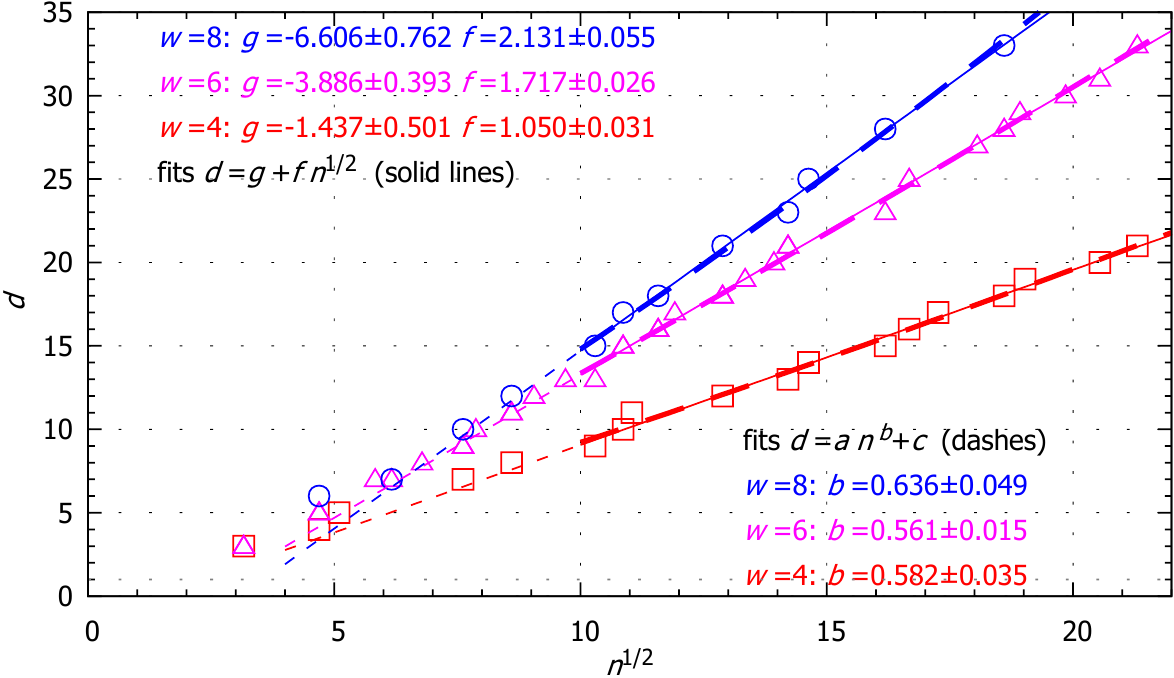}
  \caption{(Color online) Distance $d$ plotted as a function of the
    square root of the block length $n$ for a family of GB codes
    encoding $k=2$ qubits.  Squares, triangles, and circles correspond
    to row weight $w=4$, $6$, and $8$, respectively.  The fits to
    $d=g+f n^{1/2}$ using only the data with $n^{1/2}>10$ are shown
    with thin solid lines; the corresponding coefficients are given in
    the upper-left inset.  Thin dashed lines in the range
    $n^{1/2}<10$ are the continuation of the same plots outside of the
    range used for fitting.  Thick long dashes show the
    three-parameter fits to $d=a n^b +c$; the corresponding exponents
    $b$ are shown in the lower-right inset.}
  \label{fig:root}
\end{figure}

The computed distances $d$ are plotted in Fig.~\ref{fig:root} as a
function of the square root of the code length $n$, with different
symbols and colors for GB codes of row weight $4$, $6$, and $8$, as
indicated in the figure caption.  For clarity, for each $w$, only the
codes with the smallest $n$ giving the particular distance are shown
on the plots.  As expected, for each value of $n$, optimal codes with
larger $w$ show larger distances, with the $w=8$ codes giving
approximately a factor of two distance improvement compared to codes
with $w=4$ (equivalent to square lattice surface codes), e.g.,
$d_4=13$, $d_6=21$, and $d_8=23$ for $n=202$; the actual improvement
factors are different for different values of $n$.

Visually, the data in Fig.~\ref{fig:root} do not show much curvature,
indicating distance scaling close to a square root.  This is confirmed
by fitting the data to a general three-parameter power-law form
$d=an^b+c$ (thick long dashes), and a similar two-parameter fit with a
fixed power $b=1/2$ (thin lines): the corresponding lines lie more or
less on top of each other, even though there is some upward curvature
as indicated by the fitted exponents $b$ whose values exceed $1/2$ for
all three sets of data.

We should also notice that in an attempt to capture the large-$n$
features, only the data in the range $n>100$ was used in the fits.  In
fact, the three fitted values of the exponent $b$ remain the same to
three decimal places when the distance data for codes with $n\ge25$
are included, while the square-root slope coefficients $f$ show a
minor reduction by around 5\%.

\section{Conclusion}
\label{sec:conclusion}

To summarize, we have constructed several bounds on distances of
generalized bicycle codes.
Without a weight restriction, GB codes with linear in the block length
$n$ distances and encoding a sublinear number of qubits, GB codes of
rate $1/4$ with the distance scaling as a square root of $n$, as well
as codes with other rates and the distances ${\cal O}(n/\log n)^{1/2}$
are known to exist.

More important practically are LDPC GB codes with a finite row weight
$w$.  Technically, these are zero-rate codes, since any such code is
equivalent to a code local in a finite dimension $D$, see Statement
\ref{th:upper-d-bnd-GB}.  On the other hand, compared to the QHP and
conventional toric codes, GB codes with row weights $w\le 8$ may have
a factor-of-two larger distances with the same block sizes.  It
remains to be seen whether the improved distances would be sufficient
to offset the increased measurement complexity (compared to the
surface codes) due to higher stabilizer generator weights and their
non-locality.

The questions remaining for future studies include further numerical
and analytical studies of GB codes encoding $k>2$ qubits.  In addition
to studying their parameters, of interest is the analysis of their
performance in the fault-tolerant setting, as larger $k$ values also
increase the redundancy for minimum-weight stabilizer generators.

Second, remains open the question of the distance scaling for GB codes
with a bounded generator weight.  More generally, while quantum LDPC
codes with power-law distance scaling higher than a square root of the
block length have been constructed, it remains unknown whether local
in a finite dimension $D>2$ codes can beat the square root distance
bound (ignoring any logarithmic corrections).

Finally, it is the regular structure of finite-weight GB codes that
makes it possible to represent them as codes local in a
$D$-dimensional space.  Perhaps other classes of matrices in the same
CSS ansatz (\ref{eq:HxHz}) based on two commuting square matrices
would produce LDPC codes with better parameters?

\begin{acknowledgments}
  L.P.P. was financially supported in part by the NSF Division of
  Physics via grants 1820939 and 2112848, and by the Government of the
  Russian Federation through the ITMO Fellowship and Professorship
  Program.
\end{acknowledgments}

\appendix
\section{Formal proofs}
\label{sec:appendix}
\subsection{The dimensions of GB and  AB codes}

This version of the proof is equivalent to the one in
Ref.~\cite{Panteleev-Kalachev-2019}; we give it for completeness.
\begin{proof}
  Let $h(x) = \gcd (a(x),b(x),x^\ell-1)$, then the ranks of the
  double-circulant matrices (\ref{eq:HxHz}) are given by
 \begin{equation}\label{eq:rankAB}
    \rank H_X=\rank H_z = \ell-\deg h(x).
 \end{equation}
 Indeed, the ranks can be computed using the column space, as the
 number of linearly independent vectors of the form
 $\alpha A+\beta B$, where $\alpha$ and $\beta$ are length-$\ell$
 $q$-ary vectors.  Using the polynomial representation, these are
 equivalent to linearly independent polynomials of the form
 $$ \alpha(x) a(x)+\beta(x) b(x)\bmod x^\ell-1.$$
 Each term in this expression contains $h(x)$ as a factor, thus there
 can be no more than $\ell-\deg h(x)$ independent linear combinations.
 Further, $\gcd (a(x),b(x),x^\ell-1)=h(x)$ implies the existence of
 polynomials $u(x)$, $v(x)$, and $w(x)$ (B\'ezout coefficients) such
 that
 $$u(x)a(x)+v(x)b(x)+w(x)(x^\ell-1)=h(x),$$ or, equivalently, 
 $$
 u(x)a(x)+v(x)b(x)=h(x)\bmod x^\ell-1.
 $$
 Multiplying by $x^m$, we get independent linear combinations for
 $0\le m<\ell-\deg h(x)$. This proves Eq.~(\ref{eq:rankAB}), so that
 the dimension of a GB code is
 $$k= n - \rank H_X - \rank H_Z = 2\deg h(x).
 $$
 In the case of AB codes, Eq.~(\ref{eq:rankAB}) gives
 $\rank H_X'=\ell$, thus $k'=\deg h(x)$.
\end{proof}

\subsection{Proof of Statement \ref{th:AB-code-d-eq}}
\label{sec:proof-AB-code-d-eq}

\begin{proof}
  Let $[u(x), v(x)]$ be an $X$-like codeword of the GB code, it
  satisfies the polynomial equation
  \begin{equation}
    a(x) u(x)+b(x)v(x)=0\bmod x^\ell-1,\label{eq:a-b-orthogonality}  
  \end{equation}
  and, in addition, in order for the codeword to be non-trivial, for
  any $\alpha(x)\in F[x]/(x^\ell-1)$,
  \begin{equation}
    {u(x)\choose v(x)}\neq
    \alpha(x) {b(x)\choose -a(x)}\bmod x^\ell-1.\label{eq:a-b-degeneracy}  
  \end{equation}
  The coefficients of Eq.~(\ref{eq:a-b-orthogonality}) can be divided
  term-by-term by $\gcd(a,b)$, which gives 
  \begin{equation}
    a_1(x) u(x)+b_1(x)v(x)=0\bmod g(x).\label{eq:a1-b1-orthogonality}      
  \end{equation}
  Indeed, if we denote $\chi(x)\equiv \gcd(a,b)$, according to
  Eq.~(\ref{eq:h-polynomial}), $\gcd(x^\ell-1,\chi)=h(x)$, so that
  $\chi(x)$ must contain $h(x)$ as a factor, $\chi(x)=\chi_1(x)h(x)$,
  where $\chi_1(x)$ is relatively prime with $g(x)$ and, therefore, must
  be invertible modulo $g(x)$.
  
  Eq.~(\ref{eq:a1-b1-orthogonality}) has a general solution
  \begin{equation}
    \label{eq:a1-b1-solution}
    {u(x)\choose v(x)}=\xi(x){b_1(x)\choose -a_1(x)}
    +g(x){i_1(x)\choose i_2(x)}\bmod x^\ell-1,
  \end{equation}
  where $\xi(x)$, $i_1(x)$, and $i_2(x)$ are arbitrary polynomials in
  $F[x]/(x^\ell-1)$.
  Now, if we take $i_1(x)=i_2(x)=0$ with
  $\xi(x)\neq0$ and $\deg \xi(x)<\deg g(x)$, we obtain exactly the set
  of pairs $[u(x), v(x)]$ which define the distance $d_Z'$ of
  the AB code.  The condition on the degree of $\xi(x)$ follows from
  the equivalent form of the orthogonality
  condition~(\ref{eq:a1-b1-orthogonality}),
  $$
  {u(x)\choose v(x)}\neq
  \alpha'(x)h(x) {b_1(x)\choose -a_1(x)}\bmod x^\ell-1.
  $$
  Similarly, if we compare Eq.~(\ref{eq:a1-b1-solution}) with the set
  of pairs which define the distance $d_X'$ of the AB code, the
  codewords are generated by the polynomials $i_1(x)$, $i_2(x)$; for a
  non-trivial vector in the AB code we must ensure that it remains
  non-zero zero with any $\xi(x)$.  Finally, notice that all vectors
  (\ref{eq:a1-b1-solution}) that \emph{can} be made zero by choosing
  $\xi(x)$ but satisfy the condition (\ref{eq:a-b-orthogonality})
  contribute to the distance $d_X'$; the distance $d_X$ is given by
  the minimum of the union of the two sets, or, equivalently,
  $d'\equiv \min(d_X',d_Z')$.
\end{proof}

\subsection{Proof of Statement \ref{th:AB-code-dZ-upper-bnd}}
\label{sec:AB-code-dZ-upper-bnd}

\begin{proof}
  Consider a vector $0\neq e(x)=i(x)g(x)$ in the code
  ${\cal C}_{g(x)}$, where we must have $\deg i(x)<\deg h(x)$.  The
  condition for $[e(x), 0]$ to be a trivial $Z$-vector
  (degenerate to zero) in the AB code $\css(H_X',H_Z)$ reads
  \begin{equation}
    {i(x)g(x)\choose 0}=\xi(x){b_1(x)\choose -a_1(x)}\bmod x^\ell-1.
    \label{eq:degeneracy-cond}  
  \end{equation}
  To analyze this expression, it is convenient to denote 
  $$
  a_2(x)=\gcd(a_1,x^\ell-1), \quad b_2=\gcd(b_1,x^\ell-1), 
  $$
  where $\gcd(a_2,b_2)=1$ since $\gcd(a_1,b_1)=1$.  The degeneracy
  condition (\ref{eq:degeneracy-cond}) then implies that $i(x)$ must
  contain a factor $h(x)/\gcd\biglb(h(x),a_2(x)\bigrb)$.  A similar
  condition for the other vector to be trivial gives that $i(x)$ must
  contain a factor $h(x)/\gcd\biglb(h(x),b_2(x)\bigrb)$.  These
  conditions cannot be simultanelously satisfied, as in this case
  $i(x)$ would be divisible by $h(x)$, which contradicts the
  assumption.
\end{proof}

\subsection{Proof of Statement \ref{th:bound-qc-special}}
\label{sec:bound-qc-special}
\begin{proof}
Notice that the result in case (a) also follows directly from the
bound constructed in Proposition 12 of
Ref.~\onlinecite{Galindo-Hernando-Matsumoto-2018}.

In both cases, the components of the codeword $[u(x),v(x)]$ satisfy
the equation
$$
f(x)u(x)+v(x)=\xi(x) g(x)\bmod x^\ell-1,
$$
where $\xi(x)\in \fq[x]$ is arbitrary.  Thus, in case (a),
with $u(x)=0$, non-zero $v(x)$ must have $\wgt v(x)\ge d[g]$.
Otherwise, with $u(x)\neq0$, in case (a), assuming $f(x)=f_1(x)p(x)$
with $f_1(x)$ and $x^\ell-1$ relatively prime,
$v(x)=p(x)[\xi(x)q(x)-f_1(x) u(x)]\bmod x^\ell-1$, where we used the
assumption $g(x)=p(x)q(x)$.  Then, any $v(x)\neq0$ is in the code
generated by $p(x)$ and thus $\wgt v(x)\ge d[p]$, while $u(x)$ is any
non-zero, $\wgt u(x)\ge1$.  Otherwise, if $v(x)=0$, a non-zero $u(x)$
must be in the code generated by $q(x)$, which gives
$\wgt u(x)\ge d[q]$.  The result in case (a) is obtained if we notice
$d[pq]\ge d[q]$ because of the inclusion
$\mathcal{C}_{pq}\subset \mathcal{C}_q$.

In case (b), for $u(x)\neq0$ we have, instead,
$$
v(x)=p(x)[\xi(x)q(x)-f_1(x) u(x)]-r(x)u(x)\bmod x^\ell-1.
$$
With the first term non-zero, its weight is bounded by $d[p]$, so that
the total weight satisfies
$$
\wgt(u)+\wgt(v)\ge \wgt(u)+\min\biglb(0,d[p]-\wgt (r)\wgt(u)\bigrb);
$$
taking the minimum over $\wgt(u)$ gives $d_0\ge d[p]/\wgt(r)$.
Otherwise, under assumptions we have, both $u(x)$ and $v(x)$ must be
non-zero and in the code generated by $q(x)$, which gives
$d_0\ge 2d[q]$.
\end{proof}
\subsection{Proof of Statement \ref{th:gv-like-bnd}}

\begin{proof}
Consider $e=[u(x), v(x)]$ of weight $s<d_g$ with $u(x)$
non-zero.  In order for it to be a non-trivial codeword in
$\gb(hf,h)$, we need
\begin{equation*}
\begin{split}
hfu+hv=0\bmod x^\ell -1, \;\text{ and }\; \\
{u(x)\choose v(x)}\neq
\xi(x){h(x)\choose h(x)f(x)}\bmod x^\ell -1.
\end{split}
\end{equation*}
The first statement is equivalent to $fu+v=0\bmod g$.  Condition on
the weight implies that $u$ cannot be a factor of $g$; with $g$
irreducible it further implies  that $\gcd(u,g)=1$. In this case we
can find unique solution  $f=v(x)/u(x)\bmod g(x)$.
Indeed, $\gcd(u,g)=1$ implies existence of polynomials $A$, $B$ such
that $Au+Bg=1$.  Thus, starting from $fu+v=wg$ with some $w$, we have
$$
A(fu+v)+Bg=Awg+Bg,\quad u+Av=0\bmod g. 
$$
With $u\neq0$, there is exactly one polynomial $f$ with
$\deg f<m-\deg h$ in this class.  On the other hand, if $u=0$, the
condition reads $v=0\bmod g$, which is impossible since it contradicts
the assumption $s<d_g$.  Now, the number of errors $e=[u(x), v(x)]$ of
weight $s$ and $u\neq0$ is ${2m\choose s}-{m\choose s}$.  Inequality
(\ref{eq:gv-sum}) is a greedy bound that implies the existence of a
polynomial $f$ of degree smaller than $\ell -\deg h$ such that the code
$\gb(hf,h)$ contains no non-trivial codewords of weight up to $y$.
\end{proof}

\bibliography{qc_all,more_qc,lpp,ldpc,spin,rwrefs}

\end{document}